\documentclass[11pt]{amsart}
\usepackage{graphicx}
\usepackage{amssymb}
\begin{document}

\centerline{\huge\bf The Mathematical Foundation of}

\medskip
\centerline{\huge\bf Post-Quantum Cryptography}

\bigskip
\centerline{\large Chuanming Zong}

\vspace{0.8cm}
\centerline{\begin{minipage}{13cm}
{\bf Abstract.} On July 5, 2022, the National Institute of Standards and Technology announced four possible post-quantum cryptography standards, three of them are based on lattice theory and the other one is based on Hash function. It is well-known that the security of the lattice cryptography relies on the hardness of the shortest vector problem (SVP) and the closest vector problem (CVP). In fact, the SVP is a sphere packing problem and the CVP is a sphere covering problem. Furthermore, both SVP and CVP are equivalent to arithmetic problems of positive definite quadratic forms. This paper will briefly introduce the post-quantum cryptography and show its connections with sphere packing, sphere covering, and positive definite quadratic forms.
\end{minipage}}


\vspace{0.8cm}
\noindent
{\Large\bf 1. Mathematical Cryptography}

\bigskip
\noindent
In 1976, W. Diffie and M. E. Hellman \cite{Diff76} set the definition and principle of public key cryptography. Two years later, the RSA public key cryptosystem was invented by R. L. Rivest, A. Shamir and L. Adleman \cite{Rive78}. These events not only inaugurated a new era in secret communications, but also marked the birth of mathematical cryptography\footnote{Mathematical cryptography here means the public key cryptography based on mathematical theories, rather than the symmetric ciphers based on mathematical techniques.}. Since then, several other mathematical cryptosystems have been successively discovered, including the Elgamal cryptosystem, the elliptic curve cryptosystem, the Ajtai-Dwork cryptosystem, the GGH cryptosystem, the NTRU cryptosystem, and the LWE cryptosystem. In the past half century, mathematical cryptography (public key cryptography) has played a crucial role in the modern technology of computer and internet. At the same time, it has been developed into an active interdisciplinary research field between mathematics and cryptography (see \cite{Gold02,Hoff14}).

Before the Diffie-Hellman\footnote{The history of secret communication is complicated, since part of the history was also secret. For a professional introduction, we refer to the first chapter of Hoffstein, Pipher and Silverman's book.}, both the enciphering process and the deciphering process of any secret communication used the same secret key. Ciphers of this sort are known as symmetric ciphers. Assume that Bob wants to send a secret message ${\bf m}$ to Alice, they have to share a secret key ${\bf k}$. Bob first scrambles his message ${\bf m}$ by the key ${\bf k}$ to a ciphertext ${\bf c}$ and then sends it through some channel to Alice. When Alice receives the ciphertext ${\bf c}$, she uses the secret key ${\bf k}$ to unscramble it and reconstitute ${\bf m}$. During this process, if the communication channel is not secure, their adversary Eve can intercept not only the ciphertext ${\bf c}$ but also the secret key ${\bf k}$ and then reconstitute their secret message ${\bf m}$.

\medskip
\noindent
{\bf Public Key Cryptography.} In 1970s, while computers and network becoming part of everyone's daily life, symmetric ciphers were no longer efficient enough, in particular in key distribution, key management and digital signatures. In Diffie and Hellman's ideal public key cryptosystem, enciphering and deciphering are governed by distinct keys, ${\bf k}_e$ and ${\bf k}_d$, such that computing ${\bf k}_d$ from ${\bf k}_e$ is computationally infeasible. Thus, each user of the network can place his enciphering key in a public directory and each one sends messages to the other enciphered in the receiver's public enciphering key and deciphers the messages he receives using his own secret deciphering key. Let $\mathcal{K}$, $\mathcal{M}$ and $\mathcal{C}$ denote the spaces of keys, plaintexts, and ciphertexts, respectively. A key ${\bf k}\in \mathcal{K}$ is in fact a pair of keys, ${\bf k}=({\bf k}_e, {\bf k}_d)$, where ${\bf k}_e$ is the enciphering key (public key) and ${\bf k}_d$ is the deciphering key (private key). Then, the principle of the public key cryptography can be formulated as following: For each enciphering key ${\bf k}_e$ there is an encryption function
$$f_e:\ \mathcal{M}\rightarrow \mathcal{C},$$
and for each deciphering key ${\bf k}_d$ there is a decryption function
$$f_d:\ \mathcal{C}\rightarrow \mathcal{M}.$$
If ${\bf k}=({\bf k}_e, {\bf k}_d)\in \mathcal{K}$, then
$$f_d(f_e({\bf m}))={\bf m}$$
hold for all ${\bf m}\in \mathcal{M}.$ Diffie and Hellman \cite{Diff76} were not able to create such a cryptosystem. However, their great idea changes cryptography from an ancient art into a modern science\footnote{Cryptographers think that Shannon's work in 1949 on perfect secrecy marked the turning point that cryptography changed from an art to a science.}.

The public key distribution systems also offer a different approach to eliminating the need for a secure key distribution channel. In such a system, two users who wish to exchange a key communicate back and forth until they arrive at a key in common. A third party eavesdropping on this exchange
must find it computationally infeasible to compute the key from the information overheard. Let $p$ be a large prime and ${\bf g}$ be a nonzero element of $\mathbb{F}_p$ such that its order is also a large prime. Both Alice and Bob agree on $p$ and ${\bf g}$ and even make them public. First, Alice chooses an integer $\alpha$ that she keeps secret, computes
$${\bf a}\equiv {\bf g}^\alpha\quad (\rm mod\ p)$$
and sends ${\bf a}$ to Bob. At the same time, Bob chooses an integer $\beta$ that he does not reveal to anyone, computes
$${\bf b}\equiv {\bf g}^\beta\quad (\rm mod\ p)$$
and sends ${\bf b}$ to Alice. Then, Alice uses her secret integer to compute
$${\bf k}\equiv {\bf b}^\alpha\quad (\rm mod\ p)$$
and Bob uses his secret integer to compute
$${\bf k}'\equiv {\bf a}^\beta\quad (\rm mod\ p).$$
In fact, we have
$${\bf k}\equiv {\bf b}^\alpha\equiv {\bf g}^{\beta\alpha}\equiv {\bf g}^{\alpha\beta}\equiv {\bf a}^\beta\equiv {\bf k}' \quad (\rm mod\ p).$$
The common value is their exchanged key.

In this process, if the communication channel is insecure, their adversary Eve can intercept both ${\bf a}$ and ${\bf b}$. However, since it is hard to compute the value of ${\bf g}^{\alpha\beta}\ (\rm mod\ p)$ from the known values of ${\bf g}^\alpha\ (\rm mod\ p)$ and ${\bf g}^\beta\ (\rm mod\ p)$, she can not easily get the secret key ${\bf k}$ of Alice and Bob. Let $p$ be a (large) prime, let ${\bf g}$ be a primitive root for $\mathbb{F}_p$, and let ${\bf h}$ be a nonzero element of $\mathbb{F}_p$. Usually, the problem to solve the exponent equation
$${\bf g}^x\equiv {\bf h}\quad (\rm mod\ p)$$
is call the Discrete Logarithm Problem (DLP). The solution $x$ is called the discrete logarithm of ${\bf h}$ to the base ${\bf g}$ and is denoted by $\log_{\bf g}({\bf h})$. Clearly, the security of the Diffie-Hellman key exchange relies on the computational complexity of the DLP.

\medskip
\noindent
{\bf The RSA Public Key Cryptosystem.} In 1978, R. L. Rivest, A. Shamir and L. Adleman \cite{Rive78} invented the first public key cryptosystem (RSA public key cryptosystem). First, Alice chooses two large primes $p$ and $q$, keeps them in secret, defines $N=pq$ and
$$\varphi (N)=(p-1)(q-1),$$
and chooses an enciphering exponent $e$ satisfying
$${\rm gcd}(e, \varphi (N))=1.$$
In other words, $e$ and $\varphi (N)$ have no common divisor. Then, she chooses $(N,e)$ as the public key and publishes it. Of course, both Bob and Eve can get it. Second, Bob enciphers his plaintext ${\bf m}$ by Alice's key to the following ciphertext
$${\bf c}\equiv {\bf m}^e\quad (\rm mod\ N)$$
and sends it to Alice. Third, since Alice knows $\varphi (N)=(p-1)(q-1)$, she can compute $d$ satisfying
$$ed\equiv 1 \quad (\rm mod\ \varphi (N))$$
and decipher Bob's message as
$${\bf c}^d\equiv {\bf m}^{ed} \equiv {\bf m} \quad (\rm mod\ N),$$
based on Euler's formula
$$ {\bf m}^{\varphi(N)} \equiv 1 \quad (\rm mod\ N).$$

In the RSA cryptosystem, besides Euler's formula, two other mathematical results are also crucial. First, when $p$ and $q$ are known, it is relatively easy to compute the deciphering key $d$. For example, the Euclidean algorithm takes at most $2 \log_2(\varphi (N)) + 2$ iterations
to compute ${\rm gcd}(e, \varphi (N))$, it takes only a small multiple of $\log_2(\varphi (N))$ steps to compute $d$. On the other hand, without knowledge of $p$ and $q$, to factorize the large integer $N$ is hard. There are many electronic computer algorithms to factorize large integers. However, none of them are efficient enough to break the RSA cryptosystem. The computational hardness of integer factorization is the security guarantee of the RSA cryptosystem.

\medskip
\noindent
{\bf The ElGamal Public Key Cryptosystem.} Diffie and Hellman \cite{Diff76} presented the principle of public key cryptography and the key exchange by discrete logarithm, however they were not able to discover a particular public key cryptosystem. In 1985, almost a decade later, T. ElGamal \cite{ElGa85} discovered a public key cryptosystem based on discrete logarithm. First, both Alice and Bob choose and publish a large prime $p$ and an element ${\bf g}\in \mathbb{F}_p$ of large prime order. Second, Alice chooses a private key $a\in \mathbb{F}^*_p$, computes
$${\bf a}\equiv {\bf g}^a \quad (\rm mod\ p),$$
and sends ${\bf a}$ to Bob. Third, Bob randomly chooses an element $k\in \mathbb{F}^*_p$, encrypts his plaintext ${\bf m}$ by
$${\bf c}_1\equiv {\bf g}^k \quad (\rm mod\ p)$$
and
$${\bf c}_2\equiv {\bf m}\cdot {\bf a}^k \quad (\rm mod\ p),$$
and sends the ciphertext $({\bf c}_1, {\bf c}_2)$ to Alice. Finally, Alice decrypts the ciphertext as
$$\left({\bf c}_1^a\right)^{-1}\!\cdot {\bf c}_2\equiv {\bf g}^{-ka}\!\cdot {\bf m}\cdot {\bf g}^{ka}\equiv {\bf m} \quad (\rm mod\ p).$$
This cryptosystem is known as the discrete logarithm public key cryptosystem, or the ElGamal public key cryptosystem.

Clearly, from the computational complexity point of view, to compute an exponent and an inverse in $\mathbb{F}_p$ are relatively easy, and to compute a discrete logarithm is hard. The easiness makes the cryptosystem efficient for Alice and Bob, and the hardness guarantees the security of the cryptosystem.

\medskip
\noindent
{\bf The Elliptic Curve Public Key Cryptosystem.} In both RSA cryptosystem and ElGamal cryptosystem, the group property of $\mathbb{F}_p^*$ plays an fundamental role. Therefore, to explore new public key cryptosystems, it is reasonable starting from group structures. An elliptic curve $\mathbb{E}$ over a field $\mathbb{F}$ is the set of solutions to a Weierstrass equation of the form
$$y^2=x^3+\alpha x+\beta $$
together with an extra point ${\bf o}=(o, o)$, where the constants $\alpha\in \mathbb{F}$ and $\beta\in \mathbb{F}$ must satisfy
$$4\alpha^3 + 27\beta^2 \not= 0.$$
Assume that ${\bf p}=(x_1,y_1)$ and ${\bf q}=(x_2,y_2)$ are two points of such a curve $\mathbb{E}$, we define
\begin{itemize}
\item ${\bf o}+{\bf p}={\bf p}+{\bf o}={\bf p}$.
\item If $x_1=x_2$ and $y_1=-y_2$, then ${\bf p}+{\bf q}={\bf o}$.
\item Otherwise,
$${\bf p}+{\bf q}=(\lambda^2-x_1-x_2, \lambda (x_1-x_2)-y_1),$$
where
$$\lambda=\left\{ \begin{array}{ll}
{{y_2-y_1}\over {x_2-x_1}}&  \quad \mbox{if ${\bf p}\not={\bf q}$},\\
{{3x_1^2+\alpha}\over {2y_1}}&  \quad \mbox{if ${\bf p}={\bf q}$}.
\end{array}\right.$$
\end{itemize}
It is well-known that the points of an elliptic curve is a group under this additive. In particular, when $\mathbb{F}$ is a finite field, the
elliptic curve is a finite group. Therefore it is natural to investigate public key cryptosystems based on elliptic curves $\mathbb{E}$ over finite fields.

In 1985, N. Koblitz \cite{Kobl87} and V. S. Miller \cite{Mill85} independently proposed a public key cryptosystem based on elliptic curve. In this setting, the group is writing in additive rather than multiplicative. First, Alice and Bob choose a large prime $p$, an elliptic curve $\mathbb{E}$ over $\mathbb{F}_p$, and a point ${\bf p}\in \mathbb{E}$. These parameters can be made public. Second, Alice chooses a private key $n$, computes
$${\bf q}=n\hspace{0.03cm} {\bf p}={\bf p}+{\bf p}+\ldots +{\bf p},$$
and publishes the public key ${\bf q}$. Third, Bob chooses a random element $k$ and encrypts his plaintext ${\bf m}\in \mathbb{E}$ by Alice's public key as
$${\bf c}_1=k\hspace{0.03cm} {\bf p}\in \mathbb{E},$$
$${\bf c}_2={\bf m}+k\hspace{0.03cm}{\bf q}\in \mathbb{E},$$
and sends ciphertext $({\bf c}_1, {\bf c}_2)$ to Alice. Finally, Alice decrypts the ciphertext by
$${\bf c}_2-n\hspace{0.03cm}{\bf c}_1={\bf m}+k\hspace{0.03cm}{\bf q} -nk\hspace{0.03cm}{\bf p}={\bf m}+kn\hspace{0.03cm}{\bf p} -nk\hspace{0.03cm}{\bf p}={\bf m}.$$

Similar to the discrete logarithm, if ${\bf q}=n\hspace{0.03cm}{\bf p}$, we write
$$n=\log_{\bf p}({\bf q})$$
and call it the elliptic discrete logarithm of ${\bf q}$ with respect to ${\bf p}$. It is understandable that to determine the value of $\log_{\bf p}({\bf q})$ is a hard problem. Clearly, the security of the elliptic curve cryptosystem relies on the hardness of determining the elliptic discrete logarithm.

\medskip
\noindent
{\bf Lattice Public Key Cryptography.} Assume that ${\bf a}_1$, ${\bf a}_2$, $\ldots$, ${\bf a}_n$ are linearly independent vectors in the $n$-dimensional Euclidean space $\mathbb{E}^n$. We call
$$\Lambda =\left\{z_1{\bf a}_1+z_2{\bf a}_2+\ldots +z_n{\bf a}_n:\ z_i\in \mathbb{Z}\right\}$$
an $n$-dimensional lattice and call $\{ {\bf a}_1, {\bf a}_2, \ldots, {\bf a}_n\}$ a basis of the lattice $\Lambda$. If ${\bf a}_i=(a_{i1}, a_{i2}, \ldots, a_{in})$, we define $A=(a_{ij})$ to be the corresponding $n\times n$ matrix and denote the absolute value of the determinant of $A$ by ${\rm det}(\Lambda)$. Then, the lattice can be rewritten as
$$\Lambda =\left\{{\bf z}A:\ {\bf z}\in \mathbb{Z}^n\right\}.$$
Clearly, when $n\ge 2$, an $n$-dimensional lattice has infinitely many of bases, any pair of them are connected by an unimodular matrix $U$.

Lattice is a fundamental concept in mathematics, which can be traced back to Gauss, Hermite and Minkowski. It is a finitely generated free group in algebra, a generalization of the integer systems $\mathbb{Z}$ and $\mathbb{Z}^n$ in number theory, and the most regular (periodic) discrete set in $\mathbb{E}^n$ in geometry. Although natural and simple sounding, lattices are complicated objects, in particular when the dimensions are high. In 1996, M. Ajtai studied computational complexity problems about lattices which opened a gate to lattice public key cryptography. Within two years, such public key cryptosystems were created by M. Ajtai and C. Dwork \cite{Ajta97}, O. Goldreich, S. Goldwasser and S. Halevi \cite{Gold97}, and J. Hoffstein, J. Pipher and J. H. Silverman \cite{Hoff98}, respectively.

In lattice cryptography, a basis consisting of short and nearly orthogonal vectors is called a good basis. With a good basis, one can efficiently solve some hard lattice problems. For this reason, one usually chooses a good basis as the secret key of a lattice cryptosystem and takes a bad basis (a random basis) as the corresponding public key.

\medskip
\noindent
{\bf The GGH Cryptosystem.} In 1997, O. Goldreich, S. Goldwasser and S. Halevi \cite{Gold97} invented the following cryptosystem. First, Alice chooses a good basis ${\bf a}_1$, ${\bf a}_2$, $\ldots$, ${\bf a}_n$ (private key) for a lattice $\Lambda$, chooses an $n\times n$ unimodular matrix
$U$, computes a bad basis ${\bf b}_1$, ${\bf b}_2$, $\ldots$, ${\bf b}_n$ satisfying
$$B=UA$$
and publishes the basis ${\bf b}_1$, ${\bf b}_2$, $\ldots$, ${\bf b}_n$ as the public key. Second, Bob makes his massage to an $n$-dimensional small plaintext vector ${\bf m}=(m_1, m_2, \ldots, m_n)$, chooses a random small vector ${\bf v}$, encrypts ${\bf m}$ with Alice's public key as
$${\bf c}=m_1{\bf b}_1+m_2{\bf b}_2+\ldots + m_n{\bf b}_n+{\bf v}={\bf m}B+{\bf v},$$
and sends the ciphertext ${\bf c}$ to Alice. Finally, Alice uses her private key to determine the lattice point
$${\bf d}= {\bf c}-{\bf v}={\bf m}B,$$
which is closest to ${\bf c}$, and uses the public key $B$ to compute
$${\bf d}B^{-1}={\bf m}BB^{-1}={\bf m}$$
to recover the plaintext ${\bf m}$.

The security of the GGH public key cryptosystem relies on the computational hardness to determine the closest lattice point to a given point from a bad basis of the lattice. As we will see in section 3, it is indeed a hard problem. On the contrary, if one knows a particular good basis of the lattice, she/he can efficiently determine the closest lattice point, just as Alice did.

\medskip
\noindent
{\bf The Ajtai-Dwork Cryptosystem.} Different from the previous cryptosystems, the plaintext in this system is binary. In 1997, M. Ajtai and C. Dwork \cite{Ajta97} created the following cryptosystem. Let $d$ and $M$ be two parameters satisfying $d\ge n^cM$, where $c$ is a suitable constant and $n$ is the lattice dimension. First, Alice randomly picks $n-1$ linearly independent vectors ${\bf a}_1$, ${\bf a}_2$, $\ldots $, ${\bf a}_{n-1}$ satisfying $\| {\bf a}_i\|\le M$, defines $H$ to be the hyperplane spanned by them, chooses ${\bf a}_n$ to be a random vector whose distance $d^*$ from $H$ satisfying $d\le d^*\le 2d$. For convenient, let $\Lambda^*$ denote the $(n-1)$-dimensional lattice generated by ${\bf a}_1$, ${\bf a}_2$, $\ldots $, ${\bf a}_{n-1}$, and let $\Lambda$ denote the $n$-dimensional lattice with a basis ${\bf a}_1$, ${\bf a}_2$, $\ldots $, ${\bf a}_n$. She chooses a random basis ${\bf b}^*_1$, ${\bf b}^*_2$, $\ldots $, ${\bf b}^*_{n-1}$ of $\Lambda^* $ (in fact, the norm of $H$) as the private key and chooses a random basis ${\bf b}_1$, ${\bf b}_2$, $\ldots $, ${\bf b}_n$ of $\Lambda $ as the public key. Second, Bob encrypts his binary plaintext ${\bf m}$ as following: When ${\bf m}=0$, he selects a random lattice point ${\bf p}$ of $\Lambda$ and adds a small random perturbation ${\bf v}$ to it. The perturbation ${\bf v}$ vector is chosen as the sum of $O(n)$ vectors independently and uniformly distributed in the sphere of radius $n^3M$. When ${\bf m}=1$, he simply selects a random point ${\bf q}$ in $\mathbb{E}^n$, which will be far away from the lattice with high probability. In other words,
$${\bf c}=\left\{\begin{array}{ll}
{\bf p}+{\bf v}& \mbox{if ${\bf m}=0$,}\\
{\bf q}& \mbox{if ${\bf m}=1$.}
\end{array}\right.$$
Then he sends his ciphertext to Alice. Finally, Alice decrypts the ciphertext as following: Let ${\bf u}$ denote the unit norm of $H$, the unit vector ${\bf u}$ satisfying $\langle {\bf u}, {\bf a}_n\rangle >0$ and $\langle {\bf u}, {\bf a}_i\rangle =0$ for all $i=1$, $2,$ $\ldots$, $n-1$. In fact, ${\bf u}$ is the private key. She computes
$$\gamma=\left\{\langle {\bf c},{\bf u}\rangle /d^*\right\},$$
where $\{ x\}$ denotes the fractional part of $x$, and decrypts the ciphertext ${\bf c}$ as
$${\bf m}=\left\{\begin{array}{ll}
0& \mbox{if $\gamma$ is very close to $0$ or $1$,}\\
1& \mbox{otherwise.}
\end{array}\right.$$

The security of the Ajtai-Dwork public key cryptosystem relies on the computational hardness to determine the shortest lattice vector of the lattice and probability theory. As we will see in section 3, it is indeed a hard problem.

\medskip
\noindent
{\bf The NTRU Cryptosystem.} In 1998, J. Hoffstein, J. Pipher and J. H. Silverman \cite{Hoff98} discovered the following cryptosystem. Let $N$, $p$, $q$, $d_1$ and $d_2$ to be suitable integers. Let $\mathcal{R}$, $\mathcal{R}_p$ and $\mathcal{R}_q$ be three polynomial rings defined by
$$\begin{array}{lll}
\mathcal{R}&=&\mathbb{Z}[x]/\left(x^N-1\right),\\
\mathcal{R}_p&=&\left(\mathbb{Z}/p\mathbb{Z}\right)[x]/\left(x^N-1\right),\\
\mathcal{R}_q&=&\left(\mathbb{Z}/q\mathbb{Z}\right)[x]/\left(x^N-1\right),
\end{array}$$
and let $T(d_1,d_2)$ denote the set of all polynomials in $\mathcal{R}$ which has $d_1$ coefficients equal to $1$, $d_2$ coefficients equal to $-1$, and all other coefficients equal to $0$. First, Alice and Bob choose a group of public parameters $(N,p,q,d)$ such that both $N$ and $p$ prime,
$${\rm gcd}(p,q)={\rm gcd}(N,q)=1,$$
and $q>(6d+1)p$. Second, Alice chooses ${\bf k}_1\in T(d+1,d)$ and ${\bf k}_2\in T(d,d)$ as private keys, where ${\bf k}_1$ is invertible in both $\mathcal{R}_p$ and $\mathcal{R}_q$, computes the inverse ${\bf g}_p$ of ${\bf k}_1$ in $\mathcal{R}_p$ and the inverse ${\bf g}_q$ of ${\bf k}_1$ in $\mathcal{R}_q$, computes
$${\bf h}={\bf g}_q\!\cdot\! {\bf k}_2,$$
and publishes ${\bf h}$ as the public key. Third, Bob chooses a random ${\bf r}\in T(d,d)$, encrypts his plaintext ${\bf m}\in \mathcal{R}_p$ to
$${\bf c}\equiv p\!\ {\bf r}\!\cdot\! {\bf h}+{\bf m}\quad ({\rm mod}\ q),$$
and sends the ciphertext ${\bf c}$ to Alice. Finally, when Alice receives ${\bf c}$, she computes
$${\bf m}'\equiv {\bf k}_1\!\cdot\! {\bf c} \quad ({\rm mod}\ q),$$
lifts it to ${\bf m}^*\in \mathcal{R}$,
and decrypts as
$${\bf m}\equiv {\bf g}_p\!\cdot\! {\bf m}^*\quad ({\rm mod}\ p).$$
More precisely, we have
$${\bf m}'={\bf k}_1\!\cdot\! {\bf c}\equiv  p\!\ {\bf k}_1\!\cdot\! {\bf g}_q\!\cdot\! {\bf k}_2\!\cdot\! {\bf r} +{\bf k}_1\!\cdot\! {\bf m}\equiv  p\!\ {\bf k}_2\!\cdot\! {\bf r} +{\bf k}_1\!\cdot\! {\bf m} \quad ({\rm mod}\ q).$$
Since ${\bf k}_1$, ${\bf k}_2$, ${\bf r}$ and ${\bf m}$ are polynomials of small coefficients, $p\!\ {\bf k}_2\!\cdot\! {\bf r}+{\bf k}_1\!\cdot\! {\bf m}$ has coefficients within $(-q/2,q/2)$ for proper parameters. This means that
$${\bf m}^*= p\!\ {\bf k}_2\!\cdot\! {\bf r}+{\bf k}_1\!\cdot\! {\bf m}.$$

In this algebraic formulation, the NTRU cryptsystem has nothing to do with lattice. In fact, since $\mathcal{R}$ is a $N$-dimensional lattice, it can be reformulated in lattice and its security also relies on the computational hardness to determine the shortest vector problem of the lattice.

\medskip
There are several other public key cryptosystems, such as Regev's LWE cryptosystem proposed in 2005 and Gentry's fully homomorphic cryptosystem invented in 2009. Nevertheless, we will not go further to introduce them in details, since the focus of this paper is the mathematical foundation of post-quantum cryptography. For more on mathematical cryptography, we refer to J. Hoffstein, J. Pipher and J. H. Silverman \cite{Hoff14}.

\vspace{0.8cm}
\noindent
{\Large\bf 2. Post-Quantum Cryptography}

\bigskip
\noindent
Classical computer is based on the laws of electronics. Its fundamental unit of information is the binary digit (bit) $0$ or $1$. Sequences of bits are manipulated by Boolean logic gates and a succession of gates yields a computation.

\medskip
\noindent
{\bf Quantum Turing Machine.} At the beginning of 1980s, P. Benioff, R. Feynman and D. Deutsch started investigating the possibility to create a computer based on the laws of quantum mechanics. In particular, D. Deutsch \cite{Deut85} defined quantum Turing machine and quantum circuits in 1985. The fundamental unit of information (quantum bit, qubit) in such a computer may simultaneously take on every value between $0$ and $1$ with varying possibilities. The quantum computer manipulates qubits via quantum logic gates to process computation. Since the state of the output of a quantum computer can be a coherent superposition of states corresponding to different solutions of a problem, it may allow many computations to be done simultaneously and quickly.

A qubit with two states is typically represented using ket notation, in which $|0\rangle $ denotes the $0$-state and $|1\rangle $ the $1$-state. Then the (pure) states of the system have the form
$$\alpha\hspace{0.04cm} |0\rangle +\beta\hspace{0.04cm} |1\rangle,$$
where $\alpha $ and $\beta $ are complex numbers satisfying $|\alpha |^2+|\beta |^2=1.$ In an $n$-component system, the $2^n$ basis elements are represented by $|s_i\rangle =|01\ldots 0\rangle$ consisting of $n$ zeros and ones. Then, a superposition of states of the system is
$$\sum_{i=1}^{2^n}\alpha_i\hspace{0.04cm} |s_i\rangle ,$$
where $\alpha_i$ are complex numbers satisfying $|\alpha_i|^2=1,$ and $|\alpha_i|^2$ represents the possibility of the system yields state $|s_i\rangle $. A quantum logic gate will change one superposition of states to one other superposition of states. The laws of quantum mechanics only permit unitary transformations of the state and $2$-bit transformations form the building blocks of the allowable transformations, where unitary means the conjugate transpose of the transformation matrix is equal to its inverse. For example, suppose a quantum computer is in the superposition of states
$$\mbox{${i\over \sqrt{2}}$}\hspace{0.04cm}|000\rangle +\mbox{$1\over 2$}\hspace{0.04cm}|100\rangle - \mbox{$1\over 2$}\hspace{0.04cm}|110\rangle$$
and the logic gate changes the last two bits of the state by
$$\begin{array}{c}
00\\ 01\\ 10\\ 11\end{array}\rightarrow \left(\begin{array}{rrrr}
\mbox{\small ${1\over 2}$} & \mbox{\small ${1\over 2}$} & \mbox{\small ${1\over 2}$} & \mbox{\small ${1\over 2}$}\\
\mbox{\small ${1\over 2}$} & \mbox{\small ${i\over 2}$} & \mbox{\small -${1\over 2}$} & \mbox{\small -${i\over 2}$}\\
\mbox{\small ${1\over 2}$} & \mbox{\small -${1\over 2}$} & \mbox{\small ${1\over 2}$} & \mbox{\small -${1\over 2}$}\\
\mbox{\small ${1\over 2}$} & \mbox{\small -${i\over 2}$} & \mbox{\small -${1\over 2}$} & \mbox{\small ${i\over 2}$}
\end{array}\right) \begin{array}{c}
00\\ 01\\ 10\\ 11\end{array}.$$
Then, the computer will go to the superposition of states
$$\mbox{${i\over {2\sqrt{2}}}$}\left(|000\rangle + |001\rangle +|010\rangle +|011\rangle\right)+\mbox{$1\over 2$}\hspace{0.04cm}|101\rangle + \mbox{$1\over 2$}\hspace{0.04cm}|111\rangle.$$

\medskip
\noindent
{\bf Quantum Computing.} In the early 1990s, while quantum computer was not born yet, D. Deutsch, R. Jozsa and P. Shor started to explore quantum computing. First, D. Deutsch and R. Jozsa \cite{Deut92} presents a problem that can be solved by a quantum computer with certainty in polynomial time, which is exponentially less time than any classical deterministic computer, and less time than the expected time of any classical stochastic computer. Namely, given a natural number $n$ and an oracle for a function $f:\ \mathbb{Z}_{2n}\rightarrow \mathbb{Z}_2$, find a true statement in the list:
\begin{enumerate}
\item $f$ is not a constant function;
\item The sequence $f(0)$, $f(1)$, $\ldots$, $f(2n-1)$ does not contain exactly $n$ zeros.
\end{enumerate}

Almost at the same time, P. Shor \cite{Shor94} discovers a quantum polynomial time algorithms to deal with the discrete logarithm problem and the factorization problem. A decade later, J. Proos and C. Zalka \cite{Proo03} succeeds in modifying Shor's discrete logarithm quantum algorithm for elliptic curves. In other words, if there is a functioning quantum computer, Shor's algorithms can break all the RSA cryptosystem, the ElGamal cryptosystem, and the elliptic curve cryptosystem. It is hard to introduce Shor's algorithms in a page. Nevertheless, we try to explain some of his key ideas for factoring, as an example.

Let $n$ be a large old integer. If $x$ is chosen randomly and has even order $r$ modulo $n$, since
$$\left(x^{r/2}-1\right)\left(x^{r/2}+1\right)=x^r-1\equiv 0\quad ({\rm mod}\ n),$$
both ${\rm gcd}(x^{r/2}-1, n)$ and ${\rm gcd}(x^{r/2}+1, n)$ will be factors of $n$. There is a randomized reduction from factoring to the order of an element.

Let $q=2^k$ be the power of $2$ satisfying $n^2\le q<2n^2$. For any $0\le a<q$, if
$$a=\sum_{i=0}^{k-1} \alpha_i2^i$$
is the binary representation of $a$, we define the state $|a\rangle =|\alpha_{k-1}\alpha_{k-2}\cdots \alpha_0\rangle$ and define a state transformation (the Fourier transformation)
$$|a\rangle \rightarrow {1\over {q^{1/2}}}\sum_{b=0}^{q-1}{\rm exp} (2\pi iab/q)\hspace{0.05cm} |b\rangle.$$
Let $T_q$ denote the $q\times q$ matrix whose $(a,b)$ entry $t_{a,b}$ is
$$t_{a,b}= {1\over {q^{1/2}}}\hspace{0.05cm} {\rm exp} (2\pi iab/q).$$
It is easy to show that $T_q$ is a unitary transformation.

To use quantum computing to determine the order $r$ of $x$ modulo $n$, we put the first register of the machine in the uniform superposition of states representing numbers $a\ ({\rm mod}\ q)$. This leaves the machine in state
$${1\over {q^{1/2}}}\sum_{a=0}^{q-1}|a\rangle\hspace{0.04cm} |0\rangle.$$
Second, compute $x^a\ ({\rm mod}\ n)$ in the second register and leave the machine in the state
$${1\over {q^{1/2}}}\sum_{a=0}^{q-1}|a\rangle\hspace{0.04cm} |x^a\hspace{0.06cm} ({\rm mod}\ n) \rangle.$$
Third, applying the transformation $T_q$ on the first quantum register, the machine changes to the state
$${1\over q}\sum_{a=0}^{q-1}\sum_{b=0}^{q-1}{\rm exp} (2\pi iab/q)\hspace{0.05cm} |b\rangle\hspace{0.04cm} | x^a\hspace{0.06cm} ({\rm mod}\ n)\rangle.$$
Then, mathematical computation shows that the possibility of seeing state $|b\rangle$ is relatively large if there exists a rational number ${d\over r}$ satisfying
$$\left|{b\over q}-{d\over r}\right| \le {1\over {2q}},$$
where $r$ is the order of $x$. Such a fraction ${d\over r}$ and therefore the order $r$ can be found in polynomial time by using continued fraction expansion of ${b\over q}$. This quantum algorithm is polynomial time.

\medskip
\noindent
{\bf Quantum Computer.} In 1998, the first quantum computer models were demonstrated at Oxford University and IBM's Almaden Research Center.

In 2007, D-Wave demonstrated the Orion system, a 16-qubit quantum annealing processor, running three different applications at the Computer History Museum in Mountain View, California. This marked the first public demonstration of a quantum computer. In 2011, D-Wave announced D-Wave One, operating on a 128-qubit chipset using quantum annealing to solve optimization problems.

In the following years, several companies developed gate model quantum machines, including Google, IBM, Intel and Rigetti, each with different qubit designs. Gate model quantum computers use gates similar in concept to classical computers but with vastly different logic and architecture. The quantum chip is programmed by sending microwave pulses to the qubits. Digital-to-analog and analog-to-digital conversion takes place at the quantum computer chip. For example, in 2016 IBM made a 5-qubit gate model quantum computer available in the cloud to allow scientists to experiment with gate model programming. A year later, the open source Qiskit development kit and a second machine with 16 qubits were added.  In 2018, Intel announced its Tangle Lake gate model quantum chip with a unique architecture of single-electron transistors coupled together.

By 2020, there were approximately a hundred working quantum computers worldwide.

\medskip
\noindent
{\bf Post-Quantum Cryptography.} When larger and larger quantum computers are built, cryptosystems such as RSA, ElGamal and ECC will be no longer secure, post-quantum cryptography will be critical for the future of secret communication.

In 2006, the first international workshop on post-quantum cryptography took place at the Katholieke Universiteit Leuven. Since then, post-quantum cryptography has gradually become an important research branch of Cryptography.

In 2016, the National Institute of Standards and Technology launched a global project to solicit and select a handful of new encryption algorithms with the ability to resist quantum computer attacks. Six years later, after three rounds of competition and selection, the agency announced four algorithms that will underpin its future cryptography standards. They include one algorithm for general encryption and key establishment purposes (CRYSTALS-Kyber) and another three for digital signatures (CRYSTALS-Dilithium, Falcon and Sphincs$+$).

It is well-known that all CRYSTALS-Kyber, Crystals-Dilithium and Falcon are lattice based algorithms, and Sphincs$+$ is based on Hash function\footnote{Hash function is an important branch in Cryptography. It is not public key cryptography.}. Lattice cryptography was born more or less at the same time of Shor's quantum algorithms for the discrete logarithm problem and the factorization problem. It has been explored as a key candidate for post-quantum cryptography ever since.

\vspace{0.8cm}
\noindent
{\Large\bf 3. The Shortest Vector Problem and the Closest Vector Problem}

\bigskip
\noindent
In Section 1, we introduced three lattice public key cryptosystems, the GGH cryptosystem, the Ajtai-Dwork cryptosystem, and the NTRU cryptsystem. In Section 2, we mentioned that three lattice based algorithms had been chosen as post-quantum cryptography standards, CRYSTALS-Kyber, Crystals-Dilithium, and Falcon. In fact, there are many other lattice based cryptosystems and algorithms. No matter how much different in forms, the security of all those lattice based cryptosystems and algorithms rely on the computational complexity of the following two problems:

\medskip
\noindent
{\bf The Shortest Vector Problem (SVP).} {\it Find a shortest nonzero vector in an $n$-dimensional lattice $\Lambda $, i.e., find a nonzero vector
${\bf v}\in \Lambda $ that minimizes the Euclidean norm $\| {\bf v} \|$.}

\medskip
\noindent
{\bf The Closest Vector Problem (CVP).} {\it Given a vector ${\bf w}\in \mathbb{E}^n$ that is not in $\Lambda $, find a vector ${\bf v}\in \Lambda$ that is closest to ${\bf w}$, i.e., find a vector ${\bf v}\in \Lambda $ that minimizes the Euclidean norm $\| {\bf v}-{\bf w}\|.$}

\medskip
\noindent
{\bf Complexity Theory of Classic Computer.} A Turing machine $\mathcal{M}$ runs in time $t(n)$ if, for every input string ${\bf s}$ of length $n$ over some fixed input alphabet, $\mathcal{M}({\bf s})$ halts after at most $t(n)$ steps. Efficient computation with a Turing machine means that it
halts in polynomial time in the size of the input, i.e., the Turing machine runs in time $t(n) = a + n^b$ for some constants $a$ and $b$ independent of $n$.

A decision problem is the problem of deciding whether the input string satisfies or not some specified property. The class of decision problems that can be solved by a deterministic Turing machine in polynomial time is called $\mathcal{P}$. The class of decision problem that can be solved by a nondeterministic Turing machine\footnote{A nondeterministic Turing machine is a theoretical model of computation whose governing rules specify more than one possible action in some given situations.} in polynomial time is called $\mathcal{NP}$. Clearly, we have $\mathcal{P}\subseteq \mathcal{NP}$. It is widely believed that $\mathcal{P}\not= \mathcal{NP}$, i.e., there are $\mathcal{NP}$ problems that cannot be solved in deterministic polynomial time. In fact, to prove or disprove $\mathcal{P}= \mathcal{NP}$ is a fundamental problem in both mathematics and computer science.

Let $P_1$ and $P_2$ be two decision problems consisting of strings of alphabet. A reduction from $P_1$ to $P_2$ is a polynomial time computable function $f$ such that ${\bf s}\in P_1$ if and only if $f({\bf s})\in P_2$. Clearly, if $P_1$ reduces to $P_2$ and $P_2$ can be solved in polynomial time, then also $P_1$ can be solved in polynomial time. A decision problem $P$ is $\mathcal{NP}$-hard if any other $\mathcal{NP}$ problem $Q$ reduces to $P$. If $P$ is also in $\mathcal{NP}$, then $P$ is $\mathcal{NP}$-complete. Clearly, if a problem $P$ is $\mathcal{NP}$-hard, then $P$ cannot be solved in polynomial time unless $\mathcal{P}= \mathcal{NP}$.

\medskip
\noindent
{\bf The Complexity of SVP at Classic Computer.} First, a lattice may have many shortest vectors. It is known that the integer lattice $\mathbb{Z}^n$ has $2n$ shortest vectors, the two-dimensional hexagonal lattice has six shortest vectors, the three-dimensional face-centered cubic lattice has twelve shortest vectors, the eight-dimensional $E_8$ lattice has $240$ shortest vectors, and the $24$-dimensional Leech lattice has $196560$ shortest lattice vectors. In general, an $n$-dimensional lattice $\Lambda$ has at most
$$2^{0.401n(1+o(1))}$$
shortest vectors (see Section 4). However, since the lattice based cryptography uses random lattices rather than a particular one, the following result addresses the number of the shortest vectors of a random lattice.

\medskip\noindent
{\bf Theorem 3.1 (S$\ddot{\bf o}$dergren \cite{Sode11}).} {\it In $\mathbb{E}^n$, $n\ge 2$, a random lattice has exact one pair of shortest nonzero vectors.}

\medskip
Usually, lattices are given by their bases. One may intuitively believe that the bases should contain some short lattice vector. In fact, this is far away from the truth. For example, let $\Lambda$ be the integer lattice $\mathbb{Z}^2$, let $m$ be a large integer, and define ${\bf a}_1=(1, m+1)$ and ${\bf a}_2=-(1,m)$. It can be verified that $\{ {\bf a}_1, {\bf a}_2\}$ is a basis of $\Lambda$ and
$$\| {\bf a}_1\|\ge \| {\bf a}_2\|=\sqrt{1+m^2}.$$
In other words, both vectors of a basis of $\Lambda $ can be arbitrary long. Nevertheless, the length of the shortest vectors of a lattice can be bounded in terms of its determinant. In 1891, H. Minkowski \cite{Mink91} obtained the following result about the length of the shortest lattice vector.

\medskip\noindent
{\bf Theorem 3.2.} {\it Every lattice $\Lambda $ of dimension $n$ contains a nonzero vector ${\bf v}$ satisfying}
$$\| {\bf v}\| \le \left(\sqrt{2/\pi e}+o(1)\right)\sqrt{n}\ {\rm det} (\Lambda )^{1/n}.$$

\medskip
At the beginning of 1980s, about two decades before lattice cryptography was born, people started to study the computational complexity theory of lattice. In 1981, P. van Emde Boas made the following conjecture.

\medskip\noindent
{\bf Conjecture 3.1 (van Emde Boas \cite{Boas81}).} {\it The shortest vector problem is $\mathcal{NP}$-hard.}

\medskip
In the same paper, he proved that the shortest vector problem in $L_\infty$ norm is indeed $\mathcal{NP}$-hard. However, forty years later, the Euclidean case is still open today. During this long time, people also have turned to consider randomized reduction and approximation. Unlike the deterministic reduction, the randomized reduction allows the mapping function to be computable in polynomial time by a probabilistic algorithm\footnote{A probabilistic Turing machine is a non-deterministic Turing machine that chooses between the available transitions at each point according to some probability distribution. A quantum computer is another model of computation that is inherently probabilistic.}. Therefore, the output of the reduction is only required to be correct with sufficiently high probability. In 1997, M. Ajtai proved the following theorem.

\medskip\noindent
{\bf Theorem 3.3 (Ajtai \cite{Ajta98}).} {\it The shortest vector problem is $\mathcal{NP}$-hard under randomized reduction.}

\medskip
In fact, even approximation to the shortest vector is not easy. In 1998, in his Ph.D thesis D. Micciancio extended Ajtai's theorem to: To approximate the shortest vector within a factor $\sqrt{2}$ under randomized reduction is $\mathcal{NP}$-hard. In 2005, S. Khot proved the following theorem.

\medskip\noindent
{\bf Theorem 3.4 (Khot \cite{Khot05}).} {\it To approximate the shortest vector of an $n$-dimensional lattice within any constant factor $c$ under randomized reduction is $\mathcal{NP}$-hard.}

\medskip
All Ajtai, Miccincio and Khot's works deals with general $L_p$ norms. For simplicity, we only concentrate on the Euclidean case. Afterwards, Theorem 3.4 has been further extended by I. Haviv and O. Regev.

\medskip\noindent
{\bf Remark 3.1.} In 1996, M. Ajtai \cite{Ajta96} introduced a new problem, called short integer solution problem (SIS), over random $q$-ary lattices and proved the first worst-case/average-case reduction for lattice problems, that is, under certain parameters, solving SIS over the lattice chosen at random according to a certain easily samplable distribution is at least as hard as solving approximate shortest vector problem for any lattice within some polynomial factor. This result is the key bridge which leads the shortest vector problem to cryptography application.

\medskip
\noindent
{\bf The Complexity of CVP at Classic Computer.} Let $m$ be a large integer, define ${\bf a}_1=(m,0)$, ${\bf a}_2=(0,1/m)$ and define $\Lambda $ to be the two-dimensional lattice generated by ${\bf a}_1$ and ${\bf a}_2$. Clearly, we have ${\rm det}(\Lambda )=1.$ If ${\bf w}=(m/2, 1/2m)$, one can easily deduce that the distance from ${\bf w}$ to its closest lattice point is
$$\|{\bf w}, \Lambda \|=\mbox{$1\over 2$}\sqrt{m^2+1/m^2}.$$
In other words, unlike Theorem 3.2, there is no simple upper bound for the closest vector problem just in terms of the determinant of the lattice. Assume that $\{ {\bf b}_1, {\bf b}_2, \ldots, {\bf b}_n\}$ is a basis of an $n$-dimensional lattice $\Lambda$ and let ${\bf w}$ be a point in $\mathbb{E}^n$, then we have
$$\|{\bf w}, \Lambda \|\le \mbox{$1\over 2$}\sqrt{\|{\bf b}_1\|^2+\|{\bf b}_2\|^2+\ldots +\|{\bf b}_n\|^2}.$$
However, since the length of ${\bf b}_i$ can be arbitrary long, such an upper bound is not much helpful for the closest vector problem.

In 1981, when he proposed Conjecture 3.1, P. van Emde Boas proved that the CVP is $\mathcal{NP}$-hard. On the other hand, it can be shown that the CVP is in $\mathcal{NP}$ (see \cite[p.48]{Micc02}). Thus, we have the following theorem.

\medskip\noindent
{\bf Theorem 3.5 (van Emde Boas \cite{Boas81}).} {\it The closest vector problem is $\mathcal{NP}$-complete.}

\medskip
Similar to the shortest vector problem, there are many approximation hardness results about the closest vector problem. We list two of them here.

\medskip\noindent
{\bf Theorem 3.6 (Arora, Babai, Stern and Sweedyk \cite{Aror97}).} {\it To approximate the closest vector of an $n$-dimensional lattice to a given point of $\mathbb{E}^n$ within any constant factor $c$ is $\mathcal{NP}$-hard.}

\medskip\noindent
{\bf Theorem 3.7 (Dinur, Kindler, Raz and Safra \cite{Dinu03}).} {\it To approximate the closest vector of an $n$-dimensional lattice to a given point of $\mathbb{E}^n$ within factor $n^{c/\log\log n}$, where $c$ is some absolute constant, is $\mathcal{NP}$-hard.}

\medskip
It was conjectured by L. Babai in 1986 that the shortest vector problem is not harder than the closest vector problem. In 1999, this conjecture was proved by O. Goldreich, D. Micciancio, S. Safra and J.-P. Seifert.

\medskip\noindent
{\bf Theorem 3.8 (Goldreich, Micciancio, Safra and Seifert \cite{Gold99}).} {\it There is an approximation-preserving polynomial time reduction from the shortest vector problem to the closest vector problem.}

\medskip\noindent
{\bf The Lenstra-Lenstra-Lov$\acute{\rm\bf a}$sz Algorithm.} Since every pair of bases of a lattice is connected by a unimodular matrix, when the initiative basis of the lattice is not very good, one may hope to reduce it to a good one. On the other hand, it is easy to show that, if ${\bf v}_1$ is one of the shortest vectors of the lattice, it has a basis with ${\bf v}_1$ as one of the $n$ generators. In 1801, Gauss considered the shortest vector problem in two-dimensional lattices based on these facts. His idea has been developed into the following algorithm, which is known as the generalized Gauss algorithm. The input is a basis $\{ {\bf a}_1, {\bf a}_2\}$ of a two-dimensional lattice $\Lambda $. As usually, $\lfloor x\rceil$ denotes the closest integer to $x$.

\medskip
$$\begin{array}{llllll}
     & \mbox{Loop} \\
     & \qquad \mbox{If $\| {\bf a}_2\|<\| {\bf a}_1\|$, swap ${\bf a}_1$ and ${\bf a}_2$}\\
     & \qquad \mbox{Compute $m=\left\lfloor \langle {\bf a}_1,{\bf a}_2\rangle / \| {\bf a}_1\|^2\right\rceil$}\\
     & \qquad \mbox{If $m=0$, return the basis vectors ${\bf a}_1$ and ${\bf a}_2$}\\
     & \qquad \mbox{Replace ${\bf a}_2$ with ${\bf a}_2-m{\bf a}_1$}\\
     & \mbox{Continue Loop}
\end{array}$$

\medskip
It can be shown that this algorithm terminates in polynomial time of the input and produces a basis which contains a shortest vector. However, in higher dimensions, to find a solution to the shortest vector problem turns out to be extremely hard, even approximate it. In 1982, A. K. Lenstra, H. W. Lenstra Jr. and L. Lov$\acute{\rm a}$sz \cite{LLL82} proposed an algorithm, which not only can efficiently approximate the shortest vector of a lattice, but also can approximate the closest vector.

Assume that $\{ {\bf a}_1, {\bf a}_2, \ldots , {\bf a}_n\}$ is a basis of an $n$-dimensional lattice $\Lambda $. We define the associated Gram-Schmidt orthogonal basis as
$${\bf a}_i^*={\bf a}_i-\sum_{j<i}\mu_{ij}{\bf a}_j^*, \quad {\rm where}\ \mu_{ij}={{\langle {\bf a}_i,{\bf a}_j^*\rangle}\over {\langle {\bf a}_j^*,{\bf a}_j^*\rangle}}.$$

\medskip\noindent
{\bf Definition 3.1.} A basis $\{ {\bf a}_1, {\bf a}_2, \ldots , {\bf a}_n\}$ of an $n$-dimensional lattice $\Lambda $ is called to be LLL reduced if
$$|\mu_{ij}|={{|\langle {\bf a}_i,{\bf a}_j^*\rangle |}\over {\langle {\bf a}_j^*,{\bf a}_j^*\rangle}}\le {1\over 2}\ \mbox{ for all $1\le j<i\le n$}$$
and
$$\| {\bf a}_i^* \|^2 \ge \sigma \|{\bf a}_{i-1}^*\|^2 \ \mbox{ for all $i=2, 3, \ldots, n$},$$
where
$$\sigma={1\over 4}+\left({3\over 4}\right)^{n/(n-1)}.$$

\medskip\noindent
{\bf Lemma 3.1 (Lenstra, Lenstra Jr. and Lov$\acute{\rm a}$sz \cite{LLL82}).} {\it Let $\ell (\Lambda)$ be the length of the shortest vector of an $n$-dimensional lattice $\Lambda $. If $\{ {\bf a}_1, {\bf a}_2, \ldots , {\bf a}_n\}$ is a LLL reduced basis of $\Lambda $, then we have}
$$\| {\bf a}_1\|\le (2/\sqrt{3})^n\ell (\Lambda).$$

\medskip
Then, they discovered the following algorithm, known as the LLL algorithm, to search for a LLL reduced basis of an integer lattice:

\medskip
$$\begin{array}{lllllllllll}
& \mbox{Input a basis $\{ {\bf a}_1, {\bf a}_2, \ldots, {\bf a}_n\}$ for an $n$-dimensional lattice $\Lambda $}\\
& \mbox{Set $i=2$}\\
& \mbox{Set ${\bf a}_1^*={\bf a}_1$}\\
& \mbox{Loop while $i\le n$}\\
& \mbox{\quad Loop Down $j=i-1, i-2, \ldots , 1$}\\
& \mbox{\qquad Set ${\bf a}_i:={\bf a}_i-\lfloor \mu_{ij}\rceil {\bf a}_j$}\\
& \mbox{\quad End $j$ Loop}\\
& \mbox{\quad If $\| {\bf a}_i^*\|^2\ge \sigma \| {\bf a}_{i-1}^*\|^2$}\\
& \mbox{\qquad Set $i:= i+1$}\\
& \mbox{\quad Else}\\
& \mbox{\qquad Swap ${\bf a}_{i-1}$ and ${\bf a}_i$}\\
& \mbox{\qquad Set $i:= \max\ (i-1, 2)$}\\
& \mbox{\quad End If}\\
& \mbox{End $i$ Loop}\\
& \mbox{Return LLL reduced basis $\{ {\bf a}_1, {\bf a}_2, \ldots, {\bf a}_n\}$}
\end{array}$$

\medskip\noindent
{\bf Theorem 3.9 (Lenstra, Lenstra Jr. and Lov$\acute{\rm a}$sz \cite{LLL82}).} {\it Let $\Lambda$ be an $n$-dimensional integer lattice, i.e., $\Lambda \subset \mathbb{Z}^n$. The LLL algorithm terminates in polynomial time at a LLL reduced basis. Therefore, in polynomial time one can find a lattice vector ${\bf v}\in \Lambda $ satisfying}
$$\| {\bf v}\|\le (2/\sqrt{3})^n\ell (\Lambda).$$

\medskip
When the base vectors are pairwise orthogonal, to approximate the closest vector is relatively easier. In fact, a LLL reduced basis is a relatively orthogonal one. Based on the LLL reduced basis, L. Babai \cite{Baba86} proposed an algorithm in 1986 to approximate the closest vector problem. Assume that $\{ {\bf b}_1, {\bf b}_2, \ldots , {\bf b}_n\}$ is a basis of $\Lambda$ and ${\bf w}$ is a point in $\mathbb{E}^n$.

\medskip
$$\begin{array}{lllllllllll}
& \mbox{Apply LLL to $\{ {\bf b}_1, {\bf b}_2, \ldots, {\bf b}_n\}$ to find a LLL reduced basis $\{ {\bf a}_1, {\bf a}_2, \ldots, {\bf a}_n\}$}\\
& \mbox{Write ${\bf w}=t_1{\bf a}_1+t_2{\bf a}_2+\ldots +t_n{\bf a}_n$}\\
& \mbox{Set $w_i= \lfloor t_i \rceil$ for $i=1, 2, \ldots, n$}\\
& \mbox{Return the lattice vector ${\bf v}=w_1{\bf a}_1+w_2{\bf a}_2+\ldots +w_n{\bf a}_n$}
\end{array}$$

\medskip\noindent
{\bf Theorem 3.10 (Babai \cite{Baba86}).} {\it There are polynomial time algorithms approximately solve the closest vector problem within a factor $2(2/\sqrt{3})^n$. In other words, for any ${\bf w}\in \mathbb{E}^n$ one can find a lattice vector ${\bf v}\in \Lambda $ satisfying}
$$\| {\bf w}, {\bf v}\|\le 2(2/\sqrt{3})^n\|{\bf w}, \Lambda\|.$$

\medskip\noindent
{\bf Remark 3.2.} In both Theorem 3.9 and Theorem 3.10, the approximation factors are exponential of the dimensions. During the years, many efforts have been made to improve the approximation factors, such as the BKZ algorithm proposed in 1987 by C.-P. Schnorr \cite{Schn87} and R. Kannan \cite{Kann87} (see \cite[p.43-44]{Micc02}). Nevertheless, no much essential progress has been achieved (see \cite{Hoff14,Micc02}). Essentially, all this kind of algorithms are based on various types of basis reductions, which will be introduced in Section 4 and Section 5.

\medskip\noindent
{\bf Remark 3.3.} SVP and CVP have several variants which are also useful in lattice cryptography, such as GapSVP, GapCVP, the shortest basis problem (SBP), the shortest independent vector problem (SIVP), and the shortest diagonal problem (SDP). For example, assume that $\{{\bf a}_1, {\bf a}_2, \ldots, {\bf a}_n\}$ is a basis of a lattice $\Lambda$ and $d$ is a given positive number, the GapSVP with approximation factor $\alpha (n)$ asks to decide whether $\ell (\Lambda )\le d$ or $\ell (\Lambda )>  d\!\ \alpha (n)$, the SIVP with approximation factor $\alpha (n)$ asks to produce a set of $n$ linearly independent vectors of length at most  $\alpha (n) \lambda_n(\Lambda)$, where $\lambda_n(\Lambda)$ is the $n$th successive minimum of $\Lambda$. For their definitions, we refer to \cite{Micc02}.

\medskip
\noindent
{\bf The Complexity of SVP and CVP at Quantum Computer.} Since the birth of Shor's quantum algorithms for discrete logarithms and factoring in 1994, in particular since the National Institute of Standards and Technology initiated the post-quantum cryptography competition in 2016, people have tried hard to search for efficient quantum computing algorithm for the shortest vector problem and the closest vector problem, or tried to prove that there is no such algorithm. Up to now, none of the effort is succeeded. Therefore, people have turned to believe the following conjectures:

\medskip\noindent
{\bf Conjecture 3.2.} {\it There is no polynomial time quantum algorithm which can approximate the shortest vector problem within a polynomial factor.}

\medskip\noindent
{\bf Conjecture 3.3.} {\it There is no polynomial time quantum algorithm which can approximate the closest vector problem within a polynomial factor.}

\medskip
These conjectures guarantee the security of the lattice based cryptosystems as post-quantum cryptography.

\vspace{0.8cm}
\noindent
{\Large\bf 4. Sphere Packing and Sphere Covering}

\bigskip
\noindent
{\bf The Shortest Vector Problem vs Sphere Packing.} Assume that $\Lambda$ is an $n$-dimensional lattice in $\mathbb{E}^n$, with a basis $\{{\bf a}_1, {\bf a}_2, \ldots, {\bf a}_n\}.$ Let $\ell (\Lambda )$ denote the length of the shortest nonzero vectors of $\Lambda$, take $r={1\over 2}\ell (\Lambda )$, let $\kappa (\Lambda)$ be the number of the shortest nonzero vectors in $\Lambda$, and let $B^n$ denote the unit ball centered at the origin of $\mathbb{E}^n$, it is easy to see that $rB^n+\Lambda $ is a lattice sphere packing in $\mathbb{E}^n$, in which every sphere touches $\kappa (\Lambda)$ others at their boundaries. Usually, we call $rB^n+\Lambda $ a sphere packing when the spheres are pairwise interiorly disjoint. Therefore, when a lattice is given, the length of its shortest nonzero vectors is twice of the largest radius $r$ such that $rB^n+\Lambda $ is a packing.

Let $P$ be the parallelopiped defined by
$$P=\{ \alpha_1{\bf a}_1+\alpha_2{\bf a}_2+ \ldots +\alpha_n{\bf a}_n:\ 0\le \alpha_i\le 1\}.$$
Clearly, $P+\Lambda$ is a tiling of $\mathbb{E}^n$. For convenience, we write $\omega_n={\rm vol}(B^n)$. Then the quantity
$$\delta (rB^n+\Lambda)={{{\rm vol}(B^n)r^n}\over {{\rm vol}(P)}}={{\omega_n\ell (\Lambda)^n}\over {2^n{\rm det}(\Lambda)}}$$
defines a density for the sphere packing $rB^n+\Lambda $. Then, let $\mathcal{L}_n$ denote the set of all $n$-dimensional lattices, the density $\delta^*(B^n)$ of the densest lattice packing of $B^n$ and the lattice kissing number $\kappa^*(B^n)$ are defined by
$$\delta^*(B^n)=\max_{\Lambda\in \mathcal{L}_n}\delta (rB^n+\Lambda)$$
and
$$\kappa^*(B^n)=\max_{\Lambda\in \mathcal{L}_n}\kappa (\Lambda ).$$
More generally, let $\delta (B^n)$ denote the density of the densest sphere packing in $\mathbb{E}^n$ and let $\kappa (B^n)$ denote the kissing number of $B^n$, i.e., the maximal number of nonoverlapping translates of $B^n$ all touching $B^n$ at its boundary. Clearly, we have
$$\delta^*(B^n)\le \delta(B^n)$$
and
$$\kappa^*(B^n)\le \kappa(B^n).$$

In 1594, T. Harriot discovered the face-centered cubic lattice sphere packing in $\mathbb{E}^3$ and determined that its density is $\pi/\sqrt{18}=0.74\cdots$. However, he was not able to prove that the density is the maximum. Then, he told his discovery to J. Kepler. In 1611, Kepler made the following conjecture: {\it The density of the densest sphere packing in $\mathbb{E}^3$ is $\pi/\sqrt{18}$. In other words,}
$$\delta (B^3)={\pi \over {\sqrt{18}}}.$$
In 1694, I. Newton and D. Gregory discussed the following problem: {\it Can thirteen unit balls in $\mathbb{E}^3$ be brought into contact with a fixed one}? Newton thought that the maximal number of nonoverlapping translates of $B^3$ all touching $B^3$ at its boundary is twelve. In other words, he conjectured that
$$\kappa (B^3)=12.$$
However, Gregory believed that it is possible that thirteen nonoverlapping unit balls can be brought into contact with a fixed one simultaneously.
These two natural and simple sounding problems initiated a research field known as sphere packing in mathematics.

Sphere packing, to determine or estimate the values of $\delta (B^n)$, $\delta^* (B^n)$, $\kappa (B^n)$ and $\kappa^* (B^n)$, has been studied by many great mathematicians. Nevertheless, in more than four hundred years, only handful exact results have been achieved.

{\large
$$\begin{tabular}{|c|c|c||c|c|}
\hline
{\small n}&{\small $\delta^*(B^n)$}&\mbox{${{\rm Author}\atop {\rm Date}}$}&{\small $\delta (B^n)$}&\mbox{${{\rm Author}\atop {\rm Date}}$}\\
\hline
{\small 2}&${{\pi }\over {\sqrt{12}}}$&\mbox{${{\rm Lagrange}\atop {\rm 1773}}$}&${{\pi }\over {\sqrt{12}}}$ &\mbox{${{\rm Thue}\atop {\rm 1892}}$}\\
\hline
{\small 3}&${{\pi }\over {\sqrt{18}}}$&\mbox{${{\rm Gauss}\atop {\rm 1831}}$}&${{\pi }\over {\sqrt{18}}}$ &\mbox{${{\rm Hales}\atop {\rm 2005}}$}\\
\hline
{\small 4}&${{\pi^2}\over {16}}$&\mbox{${{\rm Korkin,\ Zolotarev}\atop {\rm 1872}}$}&??&??\\
\hline
{\small 5}&${{\pi^2}\over {15\sqrt{2}}}$&\mbox{${{\rm Korkin,\ Zolotarev}\atop {\rm 1877}}$}&??&??\\
\hline
{\small 6}&${{\pi^3}\over {48\sqrt{3}}}$&\mbox{${{\rm Blichfeldt}\atop {\rm 1925}}$}&??&??\\
\hline
{\small 7}&${{\pi^3}\over {105}}$&\mbox{${{\rm Blichfeldt}\atop {\rm 1926}}$}&??&??\\
\hline
{\small 8}&${{\pi^4}\over {384}}$&\mbox{${{\rm Blichfeldt}\atop {\rm 1934}}$}&${{\pi^4}\over {384}}$&\mbox{${{\rm Viazovska}\atop {\rm 2017}}$}\\
\hline
{\small 24}&${{\pi^{12}}\over {12!}}$&\mbox{${{\rm Cohn,\ Kumar}\atop {\rm 2009}}$}&${{\pi^{12}}\over {12!}}$&\mbox{${{\rm Cohn,\ Kumar,\ Miller}\atop {\rm Radchenko,\ Viazovska,\ 2017}}$}\\
\hline
\end{tabular}$$}

\centerline{Table 4.1}

{\large
$$\begin{tabular}{|c|c|c||c|c|}
\hline
{\small n}&{\small $\kappa^*(B^n)$}&\mbox{${{\rm Author}\atop {\rm Date}}$}&{\small $\kappa (B^n)$}&\mbox{${{\rm Author}\atop {\rm Date}}$}\\
\hline
{\small 2}&{\small 6} &\mbox{\small Trivial }&{\small 6} &\mbox{\small Trivial}\\
\hline
{\small 3}&{\small 12} &\mbox{${{\rm van\ der\ Waerden}\atop {\rm Sch\ddot{u}tte,\ 1953}}$}&{\small 12} &\mbox{${{\rm van\ der\ Waerden}\atop {\rm Sch\ddot{u}tte,\ 1953}}$}\\
\hline
{\small 4}&{\small 24} &\mbox{${{\rm Watson}\atop {\rm 1971}}$}&{\small 24}&\mbox{${{\rm Musin}\atop {\rm 2008}}$}\\
\hline
{\small 5}&{\small 40} &\mbox{${{\rm Watson}\atop {\rm 1971}}$}&??&??\\
\hline
{\small 6}&{\small 72} &\mbox{${{\rm Watson}\atop {\rm 1971}}$}&??&??\\
\hline
{\small 7}&{\small 126} &\mbox{${{\rm Watson}\atop {\rm 1971}}$}&??&??\\
\hline
{\small 8}&{\small 240} &\mbox{${{\rm Watson}\atop {\rm 1971}}$}&{\small 240} &\mbox{${{\rm Odlyzko,\ Sloane}\atop {\rm Leven\check{s}tein,\ 1979}}$}\\
\hline
{\small 9}&{\small 272} &\mbox{${{\rm Watson}\atop {\rm 1971}}$}&??&??\\
\hline
{\small 24}&{\small 196560} &\mbox{${{\rm Odlyzko,\ Sloane}\atop {\rm Leven\check{s}tein,\ 1979}}$}&{\small 196560}&\mbox{${{\rm Odlyzko,\ Sloane}\atop {\rm Leven\check{s}tein,\ 1979}}$}\\
\hline
\end{tabular}$$}

\centerline{Table 4.2}

\medskip
In general dimensions, let $\zeta (n)$ be the Riemann zeta-fnction, we have
$${{(n-1)\zeta (n)}\over {2^{n-1}}}\le \delta^*(B^n)\le \delta (B^n)\le 2^{-0.599n(1+o(1))},$$
where a weaker lower bound was conjectured by Minkowski in 1905, first proved by E. Hlawka in 1943, and then improved by C. L. Siegel, H. Davenport, C. A. Rogers, W. M. Schmidt and others, the upper bound was proved by G. A. Kabatjanski and V. I. Leven$\check{\rm s}$tein in 1978. For the kissing numbers, we have
$$n^{(\log_2n-2\log_2\log_2n)}\le \kappa^*(B^n)\le \kappa (B^n)\le 2^{0.401n(1+o(1))},$$
where the lower bound can be found in Conway and Sloane \cite{Conw99} and the upper bound was discovered by G. A. Kabatjanski and V. I. Leven$\check{\rm s}$tein in 1978.

There are hundreds of papers on sphere packing, employing methods and tools from various fields of mathematics. As well, there are many fascinating open problems in sphere packing. Here we list three of them as examples.

\medskip\noindent
{\bf Problem 4.1.} Determine the asymptotic orders of $\delta^*(B^n)$ and $\delta (B^n)$, if they do exist.

\medskip\noindent
{\bf Problem 4.2.} Determine the asymptotic orders of $\kappa^*(B^n)$ and $\kappa (B^n)$, if they do exist.

\medskip\noindent
{\bf Problem 4.3.} Is there a dimension $n$ satisfying $\delta^*(B^n)\not=\delta (B^n)$?

\medskip\noindent
{\bf Remark 4.1.} It is well-known that $\kappa^*(B^9)\not=\kappa(B^9)$, where $\kappa^*(B^9)=272$ and $\kappa(B^9)\ge 306.$ For more on sphere packing, we refer to \cite{Cohn17,Conw99,Zong99}.

\medskip\noindent
{\bf Remark 4.2.} Similar to the sphere case, one can define and study lattice packing of any centrally symmetric convex body, which corresponding to the shortest vector problem in different norms.

\medskip
\noindent
{\bf The Closest Vector Problem vs Sphere Covering.} Assume that $\Lambda$ is an $n$-dimensional lattice in $\mathbb{E}^n$. For every point ${\bf x}\in \mathbb{E}^n$, we define the distance between ${\bf x}$ and its closest lattice point ${\bf v}\in \Lambda$ as $\rho ({\bf x},\Lambda)$. Then, we define
$$\rho (\Lambda )=\max_{{\bf x}\in \mathbb{E}^n}\rho ({\bf x}, \Lambda).$$
It is easy to see that $\rho(\Lambda)B^n+\Lambda$ is a covering of $\mathbb{E}^n$. In fact, $\rho (\Lambda)$ is the smallest radius $r$ such that $rB^n+\Lambda$ is a covering of $\mathbb{E}^n$. Clearly, the quantity
$$\theta (\rho (\Lambda)B^n+\Lambda)={{{\rm vol}(B^n)\rho(\Lambda)^n}\over {{\rm vol}(P)}}={{\omega_n\rho(\Lambda)^n}\over {{\rm det}(\Lambda)}}$$
defines a density for the sphere covering. Then the density $\theta^*(B^n)$ of the thinnest lattice sphere covering of $\mathbb{E}^n$ is defined by
$$\theta^*(B^n)=\min_{\Lambda\in \mathcal{L}_n}\theta (\rho (\Lambda)B^n+\Lambda).$$
Similar to the packing density case, one can define the density $\theta(B^n)$ of the thinnest sphere covering.

Sphere covering, in certain sense, is regarded as a dual concept of sphere packing. In fact, they are not much related. Sphere covering came to mathematics much later that sphere packing. Up to now, our sphere covering knowledge is much limited.

{\large
$$\begin{tabular}{|c|c|c||c|c|}
\hline
{\small n}&{\small $\theta^*(B^n)$}&\mbox{${{\rm Author}\atop {\rm Date}}$}&{\small $\theta (B^n)$}&\mbox{${{\rm Author}\atop {\rm Date}}$}\\
\hline
{\small 2}&${{2\pi }\over {3\sqrt{3}}}$&\mbox{${{\rm Kersshner}\atop {\rm 1939}}$}&${{2\pi }\over {3\sqrt{3}}}$&\mbox{${{\rm Kersshner}\atop {\rm 1939}}$}\\
\hline
{\small 3}&${{5\sqrt{5}\pi }\over {24}}$&\mbox{${{\rm Bambah}\atop {\rm 1954}}$}&??&??\\
\hline
{\small 4}&${{2\pi^2}\over {5\sqrt{5}}}$&\mbox{${{\rm Delone,\ Ryskov}\atop {\rm 1963}}$}&??&??\\
\hline
{\small 5}&${{245\sqrt{35}\pi^2}\over {3888\sqrt{3}}}$&\mbox{${{\rm Ryskov,\ Baranovskii}\atop {\rm 1975}}$}&??&??\\
\hline
\end{tabular}$$}

\centerline{Table 4.3}

\medskip
In general dimensions, there is a constant $c$ such that
$${n\over \sqrt{e^3}}\lesssim \theta (B^n)\le \theta^*(B^n)\le cn(\log_en)^{\log_2\sqrt{2\pi e}},$$
where the lower bound was achieved by H. S. M. Coxeter, L. Few and C. A. Rogers in 1959, and the upper bound was discovered by Rogers in 1959 (see Rogers \cite{Roge64}).

One may realize that there is very few concrete results on sphere covering in the past half a century, in particular comparing with sphere packing. This perhaps is some indication that the closest vector problem is harder than the shortest vector problem. It is fascinating to notice that, unlike the packing case, the thinnest lattice sphere covering in $\mathbb{E}^8$ can not be achieved by the $E_8$ lattice. At least, the $A_8^*$ lattice does provide a sphere covering with a density thinner than the $E_8$ lattice. Therefore, the following problem is important and perhaps very challenging.

\medskip\noindent
{\bf Problem 4.4.} Determine the values of $\theta^*(B^8)$ and $\theta^*(B^{24})$, and their corresponding lattices.

\medskip
\noindent
{\bf Two Bridges Connecting SVP and CVP.} In 1950, C. A. Rogers \cite{Roge50} defined and studied
$$\phi^*(B^n)=\min_{\Lambda\in \mathcal{L}_n}{{2\rho (\Lambda)}\over {\ell (\Lambda)}},$$
where $\ell (\Lambda)$ is the length of the shortest nonzero vectors of $\Lambda$ and $\rho (\Lambda)$ is the maximum distance between a point ${\bf x}\in \mathbb{E}^n$ to its closest lattice point. From the intuitive point of view, one may think that $\phi^*(B^n)$ can be arbitrary large when $n\rightarrow \infty$. Surprisingly, he proved that
$$\phi^*(B^n)\le 3$$
holds in every dimension. In 1972, via mean value techniques developed by C. A. Rogers and C. L. Siegel, G. L. Butler improved Rogers' upper bound to
$$\phi^*(B^n)\leq 2+o(1).$$

The constant $\phi^*(B^n)$ has a couple of different interpretations. For example, $\phi^*(B^n)$ is the largest number such that every lattice sphere packing $B^n+\Lambda$ has a hole in which one can put a sphere of radius $\phi^*(B^n)-1$. In 1980s, several mathematicians studied this problem from different respects. Up to now, we have the following exact results.

{\large
$$\begin{tabular}{|c|c|c|c|c|c|}
\hline
$n$ &{\small $2$} &{\small $3$} &{\small $4$} &{\small $5$} \\
\hline
$\phi^*(B^n)$ &{\small $2/\sqrt3$ }&{\small $\sqrt{5/3}$} &{\small $\sqrt{2\sqrt3}(\sqrt3-1)$} &{\small $\sqrt{3/2+\sqrt{13}/6}$}\\
\hline
\mbox{${{\rm Author}\atop {\rm Date}}$} & {\small Trivial} & \mbox{${{\rm Boroczky}\atop {\rm 1986}}$} & \mbox{${{\rm Horvath}\atop {\rm 1982}}$} & \mbox{${{\rm Horvath}\atop {\rm 1986}}$}\\
\hline
\end{tabular}$$}

\centerline{Table 4.4}

\medskip
Just like the sphere covering case, there are many open important problems about $\phi^*(B^n)$. We list two of them here as examples.

\medskip\noindent
{\bf Problem 4.5.} Determine the values of $\phi^*(B^8)$ and $\phi^*(B^{24})$, and their corresponding lattices.

\medskip\noindent
{\bf Problem 4.6.} Is there a dimension $n$ such that $\phi^*(B^n)\ge 2$ ?

\medskip
The known knowledge about the Leech lattice supports the conjecture that $\phi^*(B^{24})=\sqrt{2}$. If one can improve Butler's upper bound to $\phi^*(B^n)\le 2-c$, where $c$ is a positive constant, the Minkowski-Hlawka theorem will be improved to
$$\delta^*(B^n)\ge (2-c)^{-n}.$$
On one hand, if one can find a dimension $n$ such that $\phi^*(B^n)\ge 2$, then we will get
$$\delta^*(B^n)\not=\delta (B^n),$$
which will solve Problem 4.3. It is easy to see that $\phi^*(B^n)$ can be generalized from sphere to arbitrary centrally symmetric convex bodies. For more on $\phi^*(B^n)$ and its generalizations, we refer to Zong \cite{Zong02}. Clearly, $\phi^*(B^n)$ is a bridge connecting the shortest vector problem and the closest vector problem, both are fundamental in lattice cryptography.

\medskip
There is another important notion which is closely related to both the shortest vector problem and the closest vector problem, the Dirichlet-Voronoi cell:
$$D=\left\{ {\bf x}:\ {\bf x}\in \mathbb{E}^n,\ \mbox{$\langle {\bf x}, {\bf v}\rangle \le {1\over 2}\langle {\bf v}, {\bf v}\rangle$ for all ${\bf v}\in \Lambda\setminus \{ {\bf o}\}$}\right\}.$$
Clearly, $D$ is a centrally symmetric polytope such that $D+\Lambda$ is a tiling of $\mathbb{E}^n$. Furthermore, one can deduce that
$$\ell(\Lambda)=2\min \{ \| {\bf o}, F\|:\ \mbox{$F$ is a facet of $D$}\}$$
and
$$\rho(\Lambda)=\max \{ \| {\bf o}, {\bf v}\|:\ \mbox{${\bf v}$ is a vertex of $D$}\}.$$
In fact, a shortest vector of $\Lambda $ is $2{\bf w}$ where ${\bf w}$ is a closest point of ${\bf o}$ on the boundary of $D$; a closest vector ${\bf v}\in \Lambda $ of ${\bf x}$ is the one satisfying ${\bf x}\in D+{\bf v}$.

Let us end this section with two well-known problems about the Dirichlet-Voronoi cells of lattices.

\medskip\noindent
{\bf Problem 4.7.} When $n\ge 6$, classify all the $n$-dimensional Dirichlet-Voronoi cells of lattices, i.e., determine their geometric shapes.

\medskip\noindent
{\bf Voronoi's Conjecture.} Every parallelohedron is an imagine of some lattice Dirichlet-Voronoi cell under certain linear transformation.

\medskip\noindent
{\bf Remark 4.3.} When $n\le 5$, both Problem 4.7 and Voronoi's conjecture have been solved.

\vspace{0.8cm}
\noindent
{\Large\bf 5. Positive Definite Quadratic Forms}

\bigskip
\noindent
{\bf Lattices vs Positive Definite Quadratic Forms.} Let $\Lambda $ be a lattice with a basis $\{ {\bf a}_1,$
${\bf a}_2,$ $\ldots ,$ ${\bf a}_n\}$, where ${\bf a}_i=(a_{i1}, a_{i2},$ $\ldots ,$ $a_{in})$, and let $A$ denote the $n\times n$ matrix with entries $a_{ij}$. Then, the lattice can be expressed as
$$\Lambda =\left\{{\bf z}A:\ {\bf z}\in \mathbb{Z}^n\right\}$$
and the norms of the lattice vectors can be expressed as a positive definite quadratic form
$$F({\bf z})=\langle {\bf z}A, {\bf z}A\rangle ={\bf z}AA'{\bf z}',$$
where $A'$ and ${\bf z}'$ indicate the transposes of $A$ and ${\bf z}$, respectively. On the other hand, assume that
$$F({\bf x})=\sum_{1\le i,j\le n}c_{ij}x_ix_j={\bf x}C{\bf x}'$$
is a positive definite quadratic form of $n$ variables, where $c_{ij}=c_{ji}$ and $C$ is the symmetric matrix with entries $c_{ij}$. It is known in Algebra that there is an $n\times n$ matrix $A$ satisfying $C=AA'$. Then the quadratic form also produces a lattice
$$\Lambda =\left\{{\bf z}A:\ {\bf z}\in \mathbb{Z}^n\right\}.$$
Therefore, there is a nice correspondence between lattices and positive definite quadratic forms.

\medskip
\noindent
{\bf SVP in Positive Definite Quadratic Forms.} In fact, for a lattice vector ${\bf v}={\bf z}A\in \Lambda$, we have
$$\| {\bf v}\|=\| {\bf z}A\|=\sqrt{F({\bf z})}.$$
Therefore, the shortest vector problem is equivalent to the following problem.

\medskip
\noindent
{\bf SVP in Quadratic Forms.} {\it Find an integer minimum solution for a positive definite quadratic form $F({\bf z})$, i.e., find a nonzero vector ${\bf z}\in \mathbb{Z}^n$ that minimizes the positive definite quadratic form $F({\bf z})$.}

\medskip
Let ${\rm dis} (F)$ be the discriminant of the quadratic form $F({\bf x})$ and let $\mathcal{F}_n$ denote the set of all positive definite quadratic forms of $n$ variables. Then we define
$$m(F)=\min_{{\bf z}\in \mathbb{Z}^n\setminus \{ {\bf o}\}} F({\bf z})$$
and
$$\gamma_n=\sup_{F\in \mathcal{F}_n} {{m(F)}\over {\sqrt[n]{ {\rm dis}(F)}}}.$$
Usually, $\gamma_n$ is called Hermite's constant. These constants are closely related to the densities of the densest lattice sphere packings $\delta^*(B^n)$. Since $\ell (\Lambda )=\sqrt{m(F)}$ and ${\rm dis}(F)={\rm det}(\Lambda )^2$, one can easily deduced
$$\delta^*(B^n)={{\omega_n\gamma_n^{n/2}}\over {2^n}},$$
where $\omega_n$ is the volume of the $n$-dimensional unit ball $B^n$. In fact, all the known exact results about $\delta^*(B^n)$ listed in Table 4.1 (except $\delta^*(B^{24})$) were deduced from $\gamma_n$.

{\large
$$\begin{tabular}{|c|c|c|c|c|c|c|c|c|c|}
\hline
$n$ &{\small $2$} &{\small $3$} &{\small $4$} &{\small $5$} &{\small $6$} &{\small $7$} &{\small $8$}&{\small $24$}\\
\hline
$\gamma_n$ &{\small $2/\sqrt3$ }&{\small $\sqrt[3]{2}$} &{\small $\sqrt{2}$} &{\small $\sqrt[5]{8}$} &{\small $\sqrt[6]{64\over 3}$}&{\small $\sqrt[7]{64}$} &{\small $2$} &{\small $4$}\\
\hline
\mbox{${{\rm Author}\atop {\rm Date}}$} & \mbox{${{\rm Lagrange}\atop {\rm 1773}}$} & \mbox{${{\rm Gauss}\atop {\rm 1831}}$} & \mbox{${{\rm  Zolotarev,}\atop {\rm Korkin,\ 1872}}$} & \mbox{${{\rm Zolotarev,}\atop {\rm Korkin,\ 1877}}$}&\mbox{${{\rm Blichfeldt}\atop {\rm 1925}}$}&\mbox{${{\rm Blichfeldt}\atop {\rm 1926}}$} &\mbox{${{\rm Blichfeldt}\atop {\rm 1934}}$}&\mbox{${{\rm Cohn,\ Kumar}\atop {\rm 2009}}$}\\
\hline
\end{tabular}$$}

\centerline{Table 5.1}

\medskip
Similarly, all the known lattice kissing numbers of spheres (except $\kappa^*(B^{24})$) listed in Table 4.2 were deduced from the maximum number of integer solutions to
$$F({\bf z})=m(F),$$
rather than from sphere packings. For this purpose, one need to study a particular type of quadratic forms, the ones which can be determined uniquely by the equations
$$F({\bf z}_i)=m(F).$$
Usually, such a quadratic form is called a perfect form.

\medskip
\noindent
{\bf CVP in Positive Definite Quadratic Forms.} Assume that $\Lambda =\{{\bf z}A:\ {\bf z}\in \mathbb{E}^n\}$ is an $n$-dimensional lattice in $\mathbb{E}^n$, where $A$ is an $n\times n$ nonsingular matrix. For any point ${\bf w}={\bf p}A\in \mathbb{E}^n$ and ${\bf v}={\bf z}A\in \Lambda$, we have
$$\| {\bf w}, {\bf v}\|=\| ({\bf z}-{\bf p})A\|=\sqrt{F({\bf z}-{\bf p})}.$$
Therefore, the closest vector problem is equivalent to the following problem.

\medskip
\noindent
{\bf CVP in Quadratic Forms.} {\it Given a vector ${\bf p}=(p_1, p_2, \ldots , p_n)\not\in \mathbb{Z}^n$ and a positive definite quadratic form $F({\bf x})$, find an integer vector ${\bf z}\in \mathbb{Z}^n$ that minimizes $F({\bf z}-{\bf p})$.}

\medskip
Let $Q$ denote the unit cube $\{(x_1, x_2, \ldots , x_n):\ 0\le x_i< 1\}$, let $\Lambda $ be the lattice corresponding to $F({\bf x})$, and define
$$\rho (F) =\sqrt{\max_{{\bf p}\in Q}\min_{{\bf z}\in \mathbb{Z}^n}F({\bf z}-{\bf p})}.$$
It can be verified that $\rho (F)$ is the smallest number $r$ such that $r B^n+\Lambda $ is a sphere covering of $\mathbb{E}^n$. Consequently, we get
$$\theta^*(B^n)=\min_{F\in \mathcal{F}_n}{{\omega_n\rho (F)^n}\over {\sqrt{{\rm dis}(F)}}}.$$
In fact, some known exact results about $\theta^*(B^n)$ listed in Table 4.3 were achieved by studying quadratic forms.

\medskip
\noindent
{\bf Reduction Theory of Positive Definite Quadratic Forms.} Assume that $\Lambda $ is an $n$-dimensional lattice with a basis $\{ {\bf a}_1, {\bf a}_2, \ldots, {\bf a}_n\}$, then every lattice vector ${\bf v}\in \Lambda$ can be uniquely expressed as
$${\bf v}=z_1{\bf a}_1+z_2{\bf a}_2+\ldots +z_n{\bf a}_n, \quad z_i\in \mathbb{Z},$$
and the corresponding positive definite quadratic form can be defined by
$$F({\bf z})=\langle {\bf v}, {\bf v}\rangle =\sum_{1\le i,j\le n}c_{ij}z_iz_j={\bf z}C{\bf z}',$$
where $c_{ij}=\langle {\bf a}_i, {\bf a}_j\rangle $ and $C$ is the $n\times n$ matrix with entries $c_{ij}$. Thus, many important properties of $\Lambda$ are encoded into the matrix $C$. For example, if $\{ {\bf a}_1, {\bf a}_2, \ldots, {\bf a}_n\}$ is a orthogonal basis, then we have
$$c_{ij}=\langle {\bf a}_i, {\bf a}_j\rangle =0, \quad i\not= j,$$
and therefore $C$ is a diagonal matrix. In this case, both SVP and CVP can be solved easily: The shortest basis vector is the shortest nonzero lattice vector of $\Lambda$; If ${\bf w}=w_1{\bf a}_1+w_2{\bf a}_2+\ldots +w_n{\bf a}_n\in \mathbb{E}^n$ is not a lattice vector, we take
$${\bf v}=\lfloor w_1\rceil {\bf a}_1+\lfloor w_2\rceil {\bf a}_2+\ldots +\lfloor w_n\rceil {\bf a}_n.$$
One can show that ${\bf v}\in \Lambda$ is a closest lattice vector of ${\bf w}$.

It is well-known that most lattices have no orthogonal bases. Nevertheless, every lattice has a relatively good basis with certain criterion. This is the philosophy of the reduction theory of positive definite quadratic forms and the foundation of many algorithms.

Let $U$ be a unimodular matrix and write
$${\widetilde{F}}({\bf z})={\bf z}UCU'{\bf z}'.$$
We say ${\widetilde{F}}({\bf z})$ is equivalent to $F({\bf z})$. Since the map ${\bf z}\to {\bf z}U$ is an automophism in $\mathbb{Z}^n$, one has
$$m( {\widetilde{F}})= m(F)$$
and
$${\rm dis}({\widetilde{F}}) ={\rm det}(UCU')={\rm dis}(F).$$
Let $\mathcal{F}$ be the subfamily of positive definite quadratic forms that are equivalent to $F({\bf x})$. Then, the family $\mathcal{F}_n$ can be represented as a union of different subfamilies $\mathcal{F}$. If in each subfamily $\mathcal{F}$ a particular form can be chosen, the problem of determining the values of $m(F)$, $\gamma_n$ and $\delta^*(B_n)$ can be simplified, as well as the corresponding shortest vector problem and closest vector problem. This is the basic idea of reduction theory.

In 1773, Lagrange proved that every positive definite binary quadratic form is equivalent to one satisfying
$$\left\{
\begin{array}{ll}
c_{11}\le c_{22},& \mbox{}\\
0\le 2c_{12}\le c_{11}.& \mbox{}
\end{array}\right.$$
In other words, every two-dimensional lattice has a basis $\{{\bf a}_1, {\bf a}_2\}$ such that the angle between ${\bf a}_1$ and ${\bf a}_2$ is at least $\pi/3$ and at most $\pi/2$. Then, one can deduce that $\gamma_2=2/\sqrt{3}$ and $\delta^*(B^2)=\pi/\sqrt{12}.$

In 1831, based on the work of Seeber, Gauss proved that every positive definite ternary quadratic form is equivalent to one satisfying
$$\left\{
\begin{array}{lllll}
c_{11}\le c_{22}\le c_{33},\\
0\le 2c_{12}\le c_{11},\\
0\le 2c_{13}\le c_{11},\\
0\le 2 |c_{23}|\le c_{22},\\
-2c_{23}\le c_{11}+c_{22}-2(c_{12}+c_{13}).
\end{array}\right.$$
In other words, every three-dimensional lattice has a basis $\{{\bf a}_1, {\bf a}_2, {\bf a}_3\}$ such that the angle between ${\bf a}_1$ and ${\bf a}_2$ is at least $\pi/3$ and at most $\pi/2$, the angle between ${\bf a}_1$ and ${\bf a}_3$ is at least $\pi/3$ and at most $\pi/2$, and the angle between ${\bf a}_2$ and ${\bf a}_3$ is at least $\pi/3$ and at most $2\pi/3$. Consequently, one can deduce that $\gamma_3=\sqrt[3]{2}$ and $\delta^*(B^3)=\pi/\sqrt{18}.$

In 1905, Minkowski generalized Lagrange, Seeber and Gauss' ideas into $n$ dimensions. As usual, we denote the greatest common divisor of $k$ integers $z_1$, $z_2$, $\ldots,$ $z_k$ by $(z_1,z_2, \ldots, z_k)$.

\medskip
\noindent
{\bf Definition 5.1.} A positive definite quadratic form $F({\bf x})= {\bf x}C{\bf x}'$ is said to be Minkowski reduced, if
$$c_{1j}\ge 0, \quad j=2, 3, \ldots, n,$$
and
$$F({\bf z})\ge c_{ii}, \quad i=1, 2, \ldots, n$$
for all integer vectors ${\bf z}=(z_1, z_2, \ldots , z_n)$ such that $(z_i, z_{i+1}, \ldots , z_n)=1$.

\medskip
Then, Minkowski proved the following theorem.

\medskip
\noindent
{\bf Theorem 5.1.} {\it Every positive definite quadratic form is equivalent to a Minkowski reduced one.}

\medskip
\noindent
{\bf Remark 5.1.} In terms of lattice, Minkowski's theorem says that every lattice $\Lambda$ has a basis $\{{\bf a}_1, {\bf a}_2, \ldots ,$ ${\bf a}_n\}$ such that
$$\left\|\sum z_j{\bf a}_j\right\| \ge \| {\bf a}_i\|$$
whenever $(z_i, z_{i+1}, \ldots, z_n)=1$. In particular, ${\bf a}_1$ is a shortest nonzero vector of $\Lambda$.

\medskip
One century ago, several great mathematicians had developed the arithmetic theory of positive definite quadratic forms, including Hermite, Korkin, Zolotarev, Minkowski and Voronoi. For example, they treated
$$\gamma (F)={{m(F)}\over {\sqrt[n]{ {\rm dis}(F)}}}$$
as a function of $F$ and studied particular types of forms.

\medskip
\noindent
{\bf Definition 5.2.} A positive definite quadratic form $F({\bf x})$ is called perfect if it is determined uniquely by the equations
$$F({\bf z}_i) =m(F).$$

\medskip
Then, Korkin and Zolotarev proved the following theorems.

\medskip
\noindent
{\bf Theorem 5.2.} {\it The Hermite constant $\gamma_n$ attains at perfect positive definite quadratic forms. In other words, if $\gamma (F)=\gamma_n$, $F({\bf x})$ must be a perfect positive definite quadratic form.}

\medskip\noindent
{\bf Theorem 5.3.} {\it Let
\begin{eqnarray*}
U_n({\bf x}) & = & \sum_{1\le i\le j\le n}x_ix_j,\quad n\ge 2,\\
&&\\
V_n({\bf x}) & = & U_n({\bf x})-x_1x_2,\quad n\ge 4,
\end{eqnarray*}
and
$$W_5({\bf x})=\sum_{i=1}^5(x_i)^2-{1\over 2}\sum_{i=2}^5x_1x_i+{1\over
2}\sum_{2\le i<j\le 4}x_ix_j-\sum_{i=2}^4x_ix_5.$$
For $n\le 5$, every perfect positive definite quadratic form $F({\bf x})$
with $m(F)=1$ is equivalent to one of the seven forms $U_2({\bf x}),$
$U_3({\bf x}),$ $U_4({\bf x}),$ $V_4({\bf x}),$ $U_5({\bf x}),$
$V_5({\bf x}),$ or $W_5({\bf x}).$}

\medskip
As consequences of these theorems, one can easily deduce that $\gamma_4=\sqrt{2}$, $\gamma_5=\sqrt[5]{8}$, $\delta^*(B^4)={{\pi^2}\over {16}}$ and $\delta^*(B^5)={{\pi^2}\over {15\sqrt{2}}}$.

\medskip
\noindent
{\bf Remark 5.2.} Perfect quadratic form is an important concept in the arithmetic theory of quadratic forms. It also plays the key role in determining the lattice kissing numbers of spheres for $4\le n\le 9$ listed in Table 4.2. The corresponding lattice of a perfect form is called a perfect lattice. We refer to Martinet \cite{Mart03} for more on this fascinating subject.

\medskip
In 1773, Korkin and Zolotarev proposed the following reduction.

\medskip\noindent
{\bf Definition 5.3.} A positive definite quadratic form
$F({\bf x})$ is said to be K-Z reduced if
$$F({\bf x}) = \sum_{i=1}^n c_i\bigg( x_i+\sum_{j=i+1}^n t_{ij}x_j
\bigg)^2,$$
where $|t_{ij}|\le {1\over 2}$ and
$$c_i=\min_{(z_i,z_{i+1}, \ldots , z_n)\in Z_{n-i+1}\setminus\{
{\bf o}\}}\bigg\{ \sum_{j=i}^nc_j\bigg( z_j+\sum_{k=j+1}^n t_{jk}
z_k\bigg)^2\bigg\} .$$

\medskip
Then, they proved the following theorem.

\medskip
\noindent
{\bf Theorem 5.4.} {\it Every positive definite quadratic form is equivalent to a K-Z reduced one.}

\medskip
Korkin and Zolotarev were not able to explore further in this direction since Zolotarev died in 1878 at the age of 31. However, in 1934 Blichfeldt succeeded in determining the values of $\gamma_6$, $\gamma_7$ and $\gamma_8$ by Korkin and Zolotarev's reduction theory. In terms of sphere packing, he proved the following theorem.

\medskip\noindent
{\bf Theorem 5.5.}
$$\delta^*(S_6)={{\pi^3}\over {48\sqrt{3}}},\quad
\delta^*(S_7)={{\pi^3}\over {105}},\quad {\it and}\quad
\delta^*(S_8)={{\pi^4}\over {384}}.$$

\medskip
Let $\{{\bf a}_1^*, {\bf a}_2^*, \ldots , {\bf a}_n^*\}$ be the Gram-Schmidt orthogonal basis associated to $\{{\bf a}_1, {\bf a}_2, \ldots , {\bf a}_n\}$ defined just above Definition 3.1. For every ${\bf v}\in \Lambda$, we define
$$\pi_i({\bf v})={\bf v}-\sum_{j=1}^i{{\langle {\bf v}\cdot {\bf a}_j^*\rangle}\over {\| {\bf a}_j^*\|^2}}{\bf a}_j^*.$$
Then, the Korkin-Zolotarev reduction can be reformulated into the following lattice version.

\medskip\noindent
{\bf Definition 5.4.} A basis $\{{\bf a}_1, {\bf a}_2, \ldots, {\bf a}_n\}$ of an $n$-dimensional lattice $\Lambda$ is called Korkin-Zolotarev reduced if it satisfies the following three conditions:
\begin{enumerate}
\item ${\bf a}_1$ is a shortest nonzero vector in $\Lambda$.
\item For $i=2, 3, \ldots, n$, the vector ${\bf a}_i$ is chosen such that $\pi_{i-1}({\bf a}_i)$ is the shortest nonzero vector in $\pi_{i-1}(\Lambda)$.
\item For all $1\le i<j\le n$, we have
$$|\langle\pi_{i-1}({\bf a}_i), \pi_{i-1}({\bf a}_j)\rangle |\le \mbox{${1\over 2}$}\|\pi_{i-1}({\bf a}_i)\|^2.$$
\end{enumerate}

\medskip
Based on this reduction, in 1987 Schnorr developed a generalization of the LLL algorithm, known as block Korkin-Zolotarev (BKZ) algorithm, to approximate the shortest vector problem (see \cite[p.43-44]{Micc02}).

Quadratic forms is a fundamental field in mathematics. Besides Lagrange, Gauss, Hermite, Korkin, Zolotarev, Minkowski, Voronoi and Delone, many modern mathematicians have made contributions to this field (see Martinet \cite{Mart03} and Zong \cite{Zong99}). Nevertheless, it is still far away from being understood. Perhaps, its fundamental hardness can illustrate its usefulness in cryptography.

\vspace{0.5cm}\noindent
{\bf Acknowledgement.} For helpful comments and suggestions, the author is grateful to professor Yanbin Pan and professor Yang Yu. This work is supported by the National Natural Science Foundation of China (NSFC12226006, NSFC11921001) and the Natural Key Research and Development Program of China (2018YFA0704701).

\vspace{0.8cm}\noindent
Chuanming Zong, Center for Applied Mathematics, Tianjin University, Tianjin 300072, P. R. China

\noindent
cmzong@tju.edu.cn

\end{document}